\documentclass[useAMS,usenatbib,usegraphicx]{mn2e} 
\usepackage[usenames]{color}
\usepackage{amssymb}

\newcommand{\be}{\begin{equation}}
\newcommand{\ee}{\end{equation}}
\newcommand{\bea}{\begin{eqnarray}}
\newcommand{\eea}{\end{eqnarray}}

\title[SDSS Clusters] 
{Comparing gravitational redshifts of SDSS galaxy clusters with
the magnification redshift enhancement of background BOSS galaxies}
\author[Jimeno et al.]
{Pablo Jimeno$^{1}$\thanks{E-mail: pablodavid.jimeno@ehu.es}, Tom Broadhurst$^{1,2}$, Jean Coupon$^{3}$, Keiichi Umetsu$^{4}$, Ruth Lazkoz$^{1}$\\
$^{1}$
Department of Theoretical Physics and History of Science, University of the Basque Country UPV-EHU, 48040 Bilbao, Spain\\
$^{2}$IKERBASQUE, Basque Foundation for Science, Alameda Urquijo, 36-5 48008 Bilbao, Spain\\
$^{3}$Astronomical Observatory of the University of Geneva, ch. d'Ecogia 16, 1290 Versoix, Switzerland\\
$^{4}$Institute of Astronomy and Astrophysics, Academia Sinica, P.O. Box 23-141, Taipei 10617, Taiwan}

\date{Draft version \today} 

\pagerange{\pageref{firstpage}--\pageref{lastpage}}

\begin{document} 
\maketitle

\label{firstpage} 
%%%%%%%%%%%%%%%%%%%%%%%%%%%%%%%%%%%%%%%%%%%%%%%%%%%%%%%%%%%%%%%%%%%%%%%%%%%%%%%
 
\begin{abstract} 
A clean measurement of the evolution of the galaxy cluster mass function can
significantly improve our understanding of cosmology 
from the rapid growth of cluster masses below $z<0.5$.
Here we examine the consistency of cluster catalogues selected from the SDSS by applying
two independent gravity-based methods using all available
spectroscopic redshifts from the DR10 release.
First, we detect a gravitational redshift related signal for 20,119 and 13,128
clusters with spectroscopic redshifts contained in the GMBCG and 
redMaPPer catalogues, respectively, at a level of $\sim$ -10\,km\,s$^{-1}$.
This we show is consistent with the magnitude expected using the richness-mass
relations provided by the literature and after applying recently
clarified relativistic and flux bias corrections. 
This signal is also consistent with the richest clusters in the larger
catalogue of Wen et al. (2012), corresponding to $M_{200m} \gtrsim 2
\times 10^{14}\,\mathrm{M}_\odot\,h^{-1}$, however we find no
significant detection of gravitational redshift signal for less
riched clusters, which may be related to bulk motions from substructure and spurious cluster
detections. 
Second, we find all three catalogues generate mass-dependent
levels of lensing magnification bias, which enhances the mean redshift of
flux-selected background galaxies from the BOSS survey. The
magnitude of this lensing effect is generally consistent with the corresponding
richness-mass relations advocated for the surveys. 
We conclude that all catalogues comprise a high proportion of reliable
clusters, and that the GMBCG and redMaPPer cluster finder algorithms
favor more relaxed clusters with a meaningful gravitational redshift
signal, as anticipated by the red-sequence colour selection of
the GMBCG and redMaPPer samples. 
\end{abstract}
%%%%%%%%%%%%%%%%%%%%%%%%%%%%%%%%%%%%%%%%%%%%%%%%%%%%%%%%%%%%%%%%%%%%%%%%%%%%%%%

%%%%%%%%%%%%%%%%%%%%%%%%%%%%%%%%%%%%%%%%%%%%%%%%%%%%%%%%%%%%%%%%%%%%%%%%%%%%%%%

\section{Introduction}
\label{sect:introduction}

The evolution of the mass function of galaxy clusters is sensitively
related to 
cosmology via the dynamically opposing effects of gravity and the
cosmological acceleration (see \citet{Huterer2013} for a thorough
review). It has been claimed that with only a few hundred massive
clusters below redshift $z \sim 0.5$, competitive constraints on
the standard cosmological model and a consistency check of the viability
of general relativity on cluster scales can be achieved 
\citep{Allen2004,Mantz2014,Rapetti2013}. In terms of the
standard model the parameters most sensitively constrained are the
normalization of the mass function, $\sigma_8$, and the cosmological
mean matter density, $\Omega_M$. These parameters can in principle be
constrained using 
the most massive clusters
\citep{Harrison2012,Waizmann2011,Watson2014}. 
Currently the
samples of clusters constructed for these purposes are X-ray or
Sunyaev-Zel'dovich (SZ) effect selected so that the masses are generally
inferred from 
indirect scalings derived from other samples of clusters
at lower redshift related to lensing and/or internal dynamics. 
Efforts are underway to obtain accurate masses of sizable samples of
massive clusters from deep multi-band lensing observations, such as the
CLASH survey \citep{Umetsu2014,Merten2014} and the ``Weighing the
Giants'' project \citep{vonderLinden2014}. These samples are a
substantial step forward in that lensing based masses are constructed
but they still rely on X-ray selection, with a significant scatter and
the potential for serious biases when inferring masses indirectly this
way \citep{Rozo2009a}. Ideally the sample selection would be best made by
selecting clusters in a volume limited way from densely sampled redshift
surveys with masses obtained by weak lensing. Large surveys with the
resolution for weak lensing work are underway: HSC \citep{Takada2010},
JPAS \citep{Benitez2014} and planned eBOSS, Large Synoptic Survey
Telescope (LSST), EUCLID, Wide-Field Infrared Survey Telescope (WFIRST),
and the Dark Energy Survey (DES), but currently no statistical sample of
clusters selected this way exists. 

The relation between richness and mass has been shown to be fraught with
systematic uncertainty \citep{Rozo2009a} related perhaps mainly to the
complexities of gas physics that may be expected to significantly
complicate the conversion of X-ray or SZ luminosities to total cluster
mass. Weak lensing mass measurements for subsamples of relaxed clusters can
help reduce the scatter in mass-observable scaling relations
\citep{vonderLinden2014}. Recent cluster weak lensing efforts with deep
Subaru observations have achieved an accuracy of sub-10\% in the overall
cluster mass calibration \citep{vonderLinden2014,Umetsu2014}, which is
currently limited by relatively small sample sizes. 

While we await the new lensing surveys, we can examine the new optically
selected samples of clusters constructed from the huge volume observed
by the SDSS and BOSS surveys for which many tens of thousands of
clusters have been painstakingly identified independently by several
groups using different cluster finding algorithms. 

Among some of the recent ones, based on SDSS data, we find the maxBCG
catalogue \citep{Koester2007}, which is based on red-sequence cluster
detection techniques and provided 13,823 clusters with photometric
redshifts (hereafter photo-$z$'s) using SDSS DR5 data. 
\citet{Szabo2011}, using an adaptive matched filter (AMF) cluster finder
\citep{Dong2008}, presented an optical catalogue of 69,173 clusters in the
redshift range $0.045\leqslant z < 0.78$, based on SDSS DR6 data. This
catalog, differing from others, did not rely on the presence of a
luminous central galaxy in order to detect and measure the properties of
the cluster, but provided a catalogue with the three brightest galaxies
associated to them. 
Using also DR6 photometric data, \citet{Wen2009} found 39,716 clusters
of galaxies below redshift $z=0.6$, identifying as clusters those groups
with more than eight $M_r\leqslant-21$ galaxies inside a determined
volume. 
\citet{Tempel2014} construct flux- and volume-limited galaxy groups
catalogues from SDSS spectroscopic data using a variable linking length
friends-of-friends (FoF) algorithm. The masses of the groups are estimated
using the velocity dispersion measurements via the virial theorem, and
although 82,458 groups are found, only around 2,000 of them have masses
above $10^{14} M_\odot$. 
The CAMIRA algorithm by \cite{Oguri2014}, based on colour prediction of
red-sequence galaxies in clusters, provides richness and photometric
redshift estimates for 71,743 clusters in the $0.1<z<0.6$ redshift range
using SDSS DR8 photometric data. 

Using three of the largest cluster catalogues produced to date, and
described in the next section, we relate the optical richness to
statistical measures of mass related observables, in particular we focus
here on the effects of gravitational redshift and gravitational
magnification. 

The effect of gravitational redshift is simply a consequence of the
reduced frequency of light observed for objects emitting from a lower
gravitational potential relative to the observer. This relativistic
effect has been advocated in terms of the change in frequency of
emission lines present in the hot cluster gas \citep{Broadhurst2000}
which may be applied to individual relaxed clusters where bulk gas
motions do not dominate. This new observational signature of clusters differs from others, in the sense that it provides a novel and unique way to test gravity, as modified gravity could lead to deeper potential wells inside clusters, and thus, stronger gravitational redshift than the one predicted by general relativity \citep{Gronke2014, Jain2013}. A statistical effect on the redshifts of member
galaxies has been claimed to be detected for optically selected stacked
cluster samples from the SDSS survey \citep{Wojtak2011,Dominguez2012,Sadeh2014}, for which the
brightest cluster galaxy (BCG) is found to lie systemically offset in
velocity relative to other member galaxies. 

The sense of this effect is
opposite to that induced by tangential motion
\citep{Zhao2013} and other effects related to galaxy kinematics
\citep{Kaiser2013}. All these effects combined induce a new asymmetry on the cross-correlation function \citep{Croft2013,Bonvin2014}, different from the well known redshift-space distortion asymmetry, as it depends not only on the absolute value of the line-of-sight separation from the center of the cluster, but also on its sign.

The magnification bias effect that we explore here is related to the
increased flux from background galaxies, which promotes galaxies
above the flux limit whilst magnifying the area of sky over which they
are detected, leading to greater depth for luminous background galaxies
\citep{Broadhurst1995}. A significant detection of this redshift
enhancement effect has been reported recently by \citet[][hereafter, CBU13]{Coupon2013}
combining SDSS clusters and lensing background galaxies from the BOSS
survey. Here we explore this effect further with the new data releases
in an enlarged sample of clusters and background galaxies, allowing new
correlations to be examined in this context. 
This effect has the advantage over weak lensing estimated from shear
to be free from the large intrinsic and instrumental shape dispersion.
It requires on the other hand a clean sample of background galaxies
with accurate spectroscopic redshifts, limited to fewer galaxies.

Individual massive clusters now routinely provide a measurement of
magnification bias, in terms of the background counts. This effect is
a projection over the integrated luminosity function described above,
which it has been shown reduces significantly the surface number
density of red background galaxies behind individual clusters
\citep{Broadhurst2005,Umetsu2011,Umetsu2012,Umetsu2013} and similar
effects are claimed for background QSO's and Lyman break galaxies
\citep{Ford2012,Hildebrandt2013}. The expansion of the sky by
magnification is found to dominate over the opposing increase from
objects promoted from lower luminosity above the flux limit. The
requirement for this is deep imaging \citep{Umetsu2014,Taylor1998}, so that this effect can be traced with sufficient numbers over several independent radial bins per cluster. \cite{Umetsu2011} have shown that this effect can significantly enhance the accuracy of lensing derived cluster masses when added to weak shear measurements.

In this study we provide these two independent gravitational
measurements for the three large stacked samples of SDSS clusters
described in section \ref{sect:data}. In section \ref{sect:grav_redshift}
we define the phase-space region in which we will measure the velocity
distribution of galaxies around clusters to identify any possible
gravitational redshift or internal motion related effects. For the magnification we examine the
mean redshift of background BOSS galaxies in section
\ref{sect:redshift_enhancement} and report our conclusion is section
\ref{sect:conclusions}. Throughout this paper we adopt the
standard cosmological parameters of a fiducial flat $\Lambda$CDM
cosmology \citep{Komatsu2011} with $H_0=72$\,km\,s$^{-1}$\,Mpc and
$\Omega_m=0.26$. To quantify cluster masses we adopt $\mathrm{M}_{200m}$ units, i.e., mass measured with respect to 200 times the \textit{mean} background density of the universe. We use the prescription given by \citet{Hu2003} to convert between different mass definitions, which assumes a Navarro-Frenk-White
(\citealt{Navarro1997}, hereafter NFW) halo density profile, using the concentration-mass relations provided by \cite{Bhattacharya2013}.

\section{Data}
\label{sect:data}

We use the data from the Sloan Digital Sky Survey (SDSS),
a combined photometric and spectroscopic survey conducted
on a 2.5-meter wide angle telescope located at Apache Point
Observatory \citep{Gunn2006}. The SDSS covers a unique
footprint of 14,555 deg$^2$ of sky, and comprises optical imaging data of
nearly 500 million unique objects in five filters ($u$,
$g$, $r$, $i$ and $z$)
and over 1,600,000 unique spectra obtained with the original SDSS spectrograph (640 spectroscopic fibers
per plate) under the Legacy programme. The ongoing BOSS programme
aims at complementing the spectroscopic sample with a total of 1,400,000 color-selected
galaxies in the range $0.3 < z< 0.7$ with the newly
installed BOSS spectrograph (1,000 fibers per plate,
\citealt{Smee2013}). 
Here we use the DR10 release which contains all galaxies with reliable
spectroscopic measurements from the Legacy programme plus more than 850,000 galaxies
from the BOSS programme.

\subsection{SDSS Cluster Catalogues}
\label{sect:cluster_cats}

\begin{figure}
\resizebox{84mm}{!}{\includegraphics{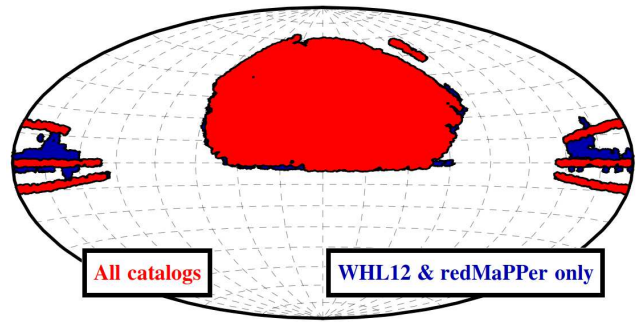}}
\caption{Clusters distribution on the sky, using equal-area map projection. We are considering here the final sample of clusters that we will be using in our study, that is, those with good spectroscopic measurements of the central BCG. This leaves a final sample of 20,119 for the GMBCG catalogue case, 52,682 for the WHL12 catalogue, and 13,128 for the redMaPPer cluster catalogue. After the restrictions applied in Sec.~\ref{sect:grav_redshift}, these samples are reduced to 4,278, 12,661 and 3,372 clusters, respectively.}
\label{fig:Clusters_Sky_Distribution}
\end{figure}

\begin{figure}
\resizebox{84mm}{!}{\includegraphics{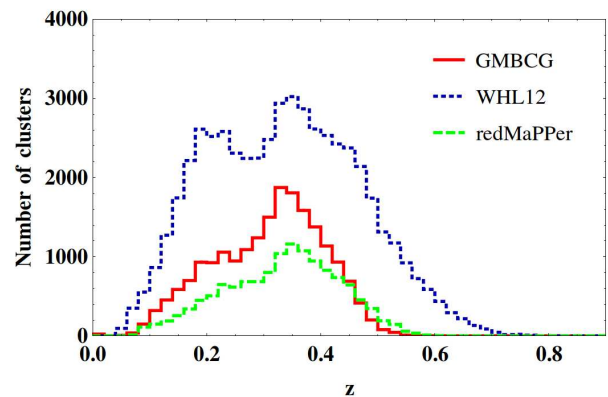}}
\caption{GMBCG (dotted, red), WHL12 (continuous, blue) and redMaPPer (dashed, green) cluster redshift distributions.}
\label{fig:Clusters_Redshift_Distribution}
\end{figure}

\begin{figure}
\resizebox{84mm}{!}{\includegraphics{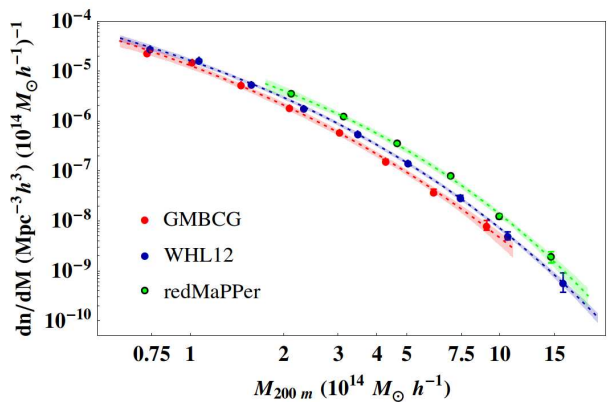}}
\caption{Observed mass distributions of clusters. Here $dn/dM$ is the differential number of clusters per unit of comoving volume and unit of mass. In order to model this mass distribution, for each cluster sample we fitted a functional mass function, shown as the dotted lines, which spans from the least massive cluster contained in each catalog, to the most massive one. The shaded regions represent 90\% confidence range.}
\label{fig:Clusters_Mass_Distribution}
\end{figure}

Large numbers of clusters with spectroscopic redshift measurements are needed to statistically investigate their gravitational redshift and lensing properties, so here we will focus our study on the three catalogues described below, which offer the largest samples useful for us to date. Throughout this paper we convert the richness observable into mass for comparison purposes, using the richness-mass relations appropriate for each case. 
The sky, redshift and mass distributions of these resulting cluster
samples and their mass distributions are shown in
Figs.~\ref{fig:Clusters_Sky_Distribution},~\ref{fig:Clusters_Redshift_Distribution},
and \ref{fig:Clusters_Mass_Distribution}. We summarize the cluster samples properties in Table \ref{tab:clusterprops}, where the final number of clusters considered in this study, N$_{\mathrm{clusters}}$, takes into account the restrictions applied in Sec.~\ref{sect:grav_redshift}.

\begin{table}
\label{symbols}
\begin{tabular}{@{}lcccc}
\hline
Catalogue&N$_{\mathrm{clusters}}$&$\langle z \rangle$&$\langle M_{200m} \rangle [10^{14}\,\mathrm{M}_\odot \, h^{-1}]$&Ref.\\
\hline
GMBCG&4,278&$0.22$&$1.5$&(1)\\
WHL12&12,661&$0.19$&$1.4$&(2)\\
redMaPPer&3,372&$0.23$&$3.2$&(3)\\
\hline
\end{tabular}
\caption{Properties of the cluster samples considered. References:
(1): \citet{Hao2010},
(2): \citet{Wen2012},
(3): \citet{Rykoff2014}.}
\label{tab:clusterprops}
\end{table}

\subsubsection{GMBCG Cluster Catalogue}

We use the optical-based cluster catalogue presented by \citet{Hao2010},
obtained applying 
the ``Gaussian Mixture Brightest Cluster Galaxy''
(GMBCG) algorithm to the SDSS DR7 data.
This cluster finding algorithm relies on the
galaxy red sequence pattern and the presence of a BCG as key features
of galaxy clusters. The SDSS photometric and redshift catalogues are
used to determine BCG candidates. To estimate the
richness $R$ of the cluster, a combination of Gaussian fitters in
colour space is used to identify overdensities around a BCG candidate
among galaxies brighter than $0.4L^*$ 
 (see \citealt{Blanton2003}), closer than 0.5\,Mpc from
the BCG, and within a photo-$z$ range of $\pm 0.25$. 
Then a circular aperture scaled to the amplitude of the overdensity
around the BCG is set to recompute the richness of the cluster. Only clusters with $ R
\geqslant 8$ are included in the final catalogue. The resulting sample
created from the application of this method comprises 55,424 clusters,
and is approximately volume limited up to redshift $z \sim 0.4$,
showing high purity and completeness in this range. Of all these
clusters we will be interested in 20,119 of them, which have
spectroscopic redshift measurements of their associated BCGs. We will
refer to this sample of clusters as the ``GMBCG'' catalogue.

We use the richness-mass relation provided by the authors of the
maxBCG catalogue (\citealp{Rozo2009a}, \citeyear{Rozo2009b}), which
uses the same richness definition as the GMBCG catalog:
\begin{equation}
M_{500c} = \mathrm{exp}(0.62) \, \left ( \frac{R}{40} \right )^{1.06}\times 10^{14}\,\mathrm{M}_\odot\,h^{-1}\,,
\end{equation}
where $R$ accounts for the richness estimation provided by the
catalog, and $M_{500c}$ is the cluster mass contained within the
radius $r_{500c}$, where the mean density of the cluster is 500 times
the \textit{critical} density of the universe at the redshift of the
cluster. We then convert $\mathrm{M}_{500c}$ into $\mathrm{M}_{200m}$.
Finally, we would like to clarify that we will be working with the richness
measurement recommended by the catalogue authors, that is, \texttt{GM\_Ngals\_weighted} instead of
\texttt{GM\_Scaled\_Ngals} when \texttt{WeightOK} is set equal to 1.

\subsubsection{WHL12 Cluster Catalogue}

Using photometric redshifts, \citet{Wen2012} 
have identified 132,684 clusters from SDSS DR8 below redshift $z \sim 0.8$. A
FoF algorithm links galaxies closer than 0.5\,Mpc in the transverse
direction and with a photo-$z$ value differing less than $\pm
0.04(1+z)$. When an overdensity is detected, the galaxy with the
maximum number of links to other cluster candidates is taken as a
temporary center, and the BCG is identified as the brightest among
those galaxies closer than a linking length from this temporary
center. Then, the total luminosity of the cluster candidate in the
$r$-band is calculated as the sum of all those members with
luminosities brighter than $0.4L^*$, and used to estimate its richness
$R_{L^*}$. A galaxy cluster is included in the catalogue if $R_{L^*}\geq
12$. Because of the magnitude limit of the SDSS photometric data, this
catalogue is claimed to be complete up to redshift $z \sim 0.42$ to a
$95 \%$ level, in the sense that there are almost no missing members
among the galaxies contributing to the estimation of the cluster
richness. Among these 132,684 clusters, 52,682 of them have
spectroscopic redshifts of the BCGs, obtained from SDSS DR9, and lie
within the region of interest for us. As we can see from Fig.~\ref{fig:Clusters_Redshift_Distribution}, WHL12 is the catalogue that provides the largest sample of clusters in all redshift ranges. From now on, this catalogue will
be referred to as ``WHL12''.

We use the richness-mass relation provided by the authors of the catalog:
\begin{equation}
M_{200c} = 10^{-1.63}\,R^{1.17}\times 10^{14}\,\mathrm{M}_\odot\,h^{-1}\,,
\end{equation}
calibrated using X-ray and lensing data. We then convert $M_{200c}$ into
$M_{200m}$.

\subsubsection{redMaPPer Cluster Catalogue}

More recently, \citet{Rykoff2014} have presented a ``red-sequence Matched-filter Probabilistic
Percolation'' (redMaPPer) cluster finding algorithm, prepared to process large amounts of photometric data. It may be
considered as an improved version of the maxBCG and the GMBCG cluster finding algorithms, as the red-sequence cluster detection and the richness estimation process have been developed using the lessons obtained from these two previous catalogs. This algorithm makes use of spectroscopic data to self-train the red-sequence model that is used to find clusters within the data. The authors of this catalogue argue that their method outperforms photo-z algorithm finders in the redshift range where this catalogue is defined, although for higher redshifts these will perform better as the red-sequence clusters are of low contrast. 

The richness estimator, $\lambda$, developed for this sample is based on the previous optical single-color richness estimator, $\lambda_\mathrm{col}$, of \cite{Rozo2009a} and \cite{Rykoff2012}, with several improvements to take into account things such as the survey mask, the probability of each galaxy to belong to the cluster, the contribution of foreground and background galaxies, etc. We refer the reader to \citet{Rykoff2014} for further details. In order to obtain a mass estimate from the richness provided by the catalog, we will use the richness-mass relation as provided in \cite{Rykoff2012}:
\begin{equation}
M_{200m} = \mathrm{exp}(1.69) \, \left ( \frac{\lambda_\mathrm{col}}{60} \right )^{1.08}\times 10^{14}\,\mathrm{M}_\odot\,h^{-1}\,.
\end{equation}
Although the richness estimators $\lambda_\mathrm{col}$ and $\lambda$ differ in many aspects, it is shown that the median deviation between them is no larger than 10\%. It may be noted that this does not provide a rigorous mass calibration, as it is based on abundance matching techniques using a \cite{Tinker2008} mass function, and it has not been corrected for selection effects. A more precise richness-mass relation is announced to be released in the future by the authors. Meanwhile, for comparison purposes, we will make use of the current relation provided.

The redMaPPer algorithm has been applied to the SDSS DR8 photometric
catalogue. In order to provide a robust cluster catalogue,
 a conservative cut of $\lambda / S(z) > 20$ (corresponding
to $M_{200m} > 1.75 \times 10^{14}\,\mathrm{M}_\odot \, h^{-1}$) is
applied to the algorithm finder, where a scaling factor $S(z)$ is
introduced to correct for the survey depth, so that for $z <
0.35$, $S(z)=1$, and the number of galaxies observed is equal to
$\lambda$, whereas it is equal to $\lambda/S(z)$ for $z \geqslant 0.35$. A total
number of 25,236 clusters were obtained in the redshift range $0.08
\leqslant z \leqslant 0.55$. Of these, we will be using 13,128, which
also have spectroscopic measurements of their BCG.

This catalogue is claimed to be volume-limited up to redshift $z \sim
0.35$ with a purity $>95\%$, where purity in this case is defined in a
way such that ``impurities'' represent richness measurements affected
by projection effects. For $\lambda>30$ and $z < 0.3$, the
completeness is as high as $\gtrsim 99\%$.

\subsection{SDSS spectroscopic samples}

We have independently used the ``Legacy'' and ``BOSS'' spectroscopic samples in the SDSS to select the cluster galaxies needed to measure the gravitational redshift effect, and the background galaxies needed for the redshift enhancement, respectively.

\begin{figure*}
 \centering
 \includegraphics[width=0.48\textwidth]{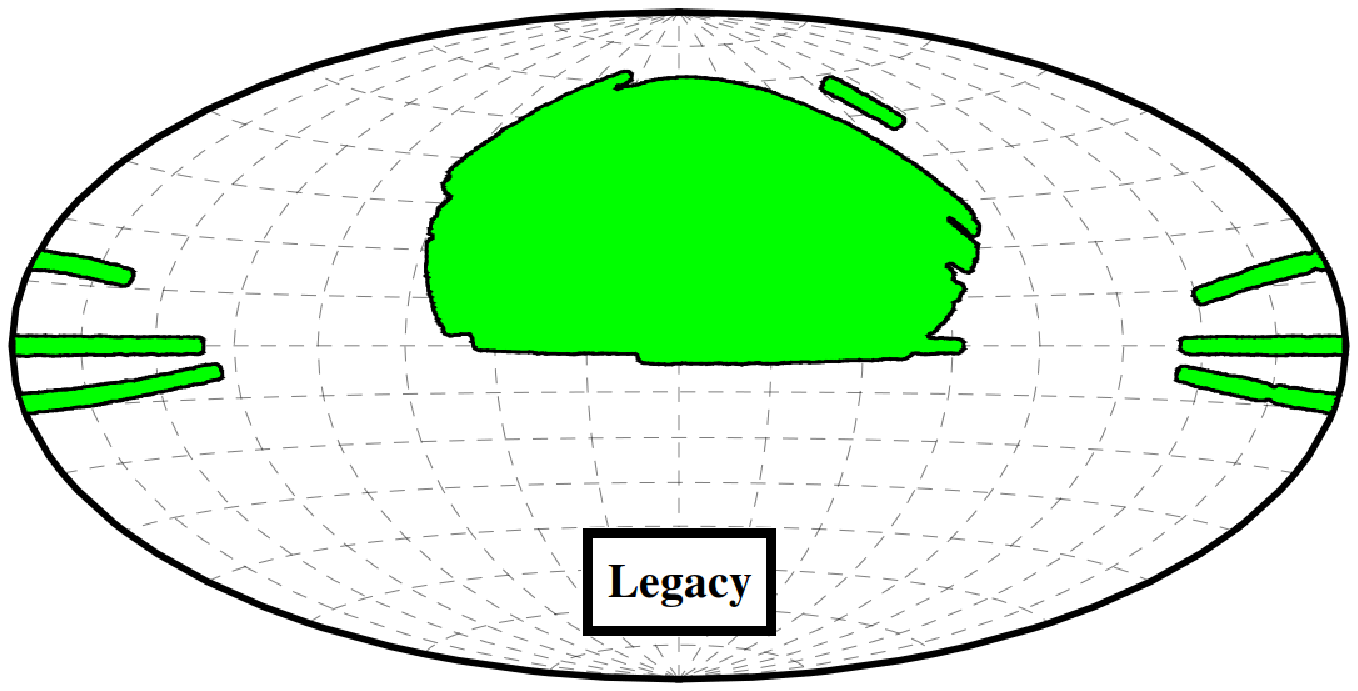}
 \includegraphics[width=0.48\textwidth]{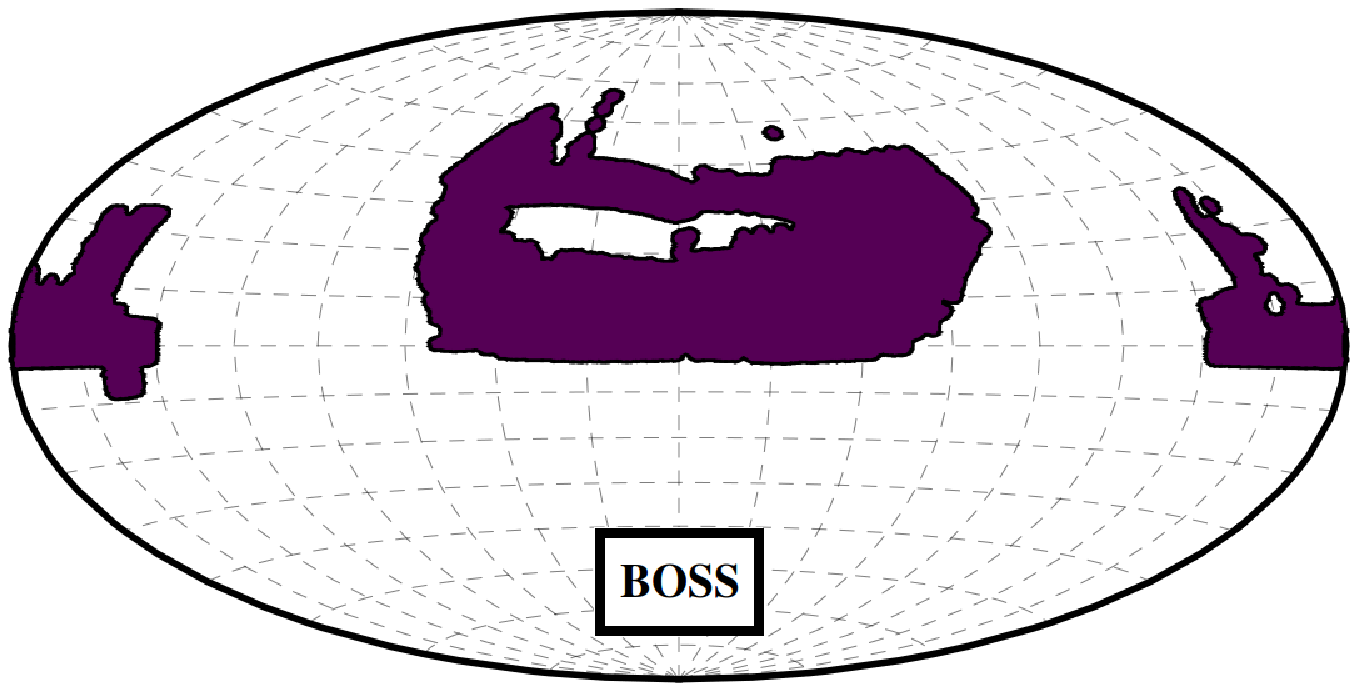}
 \caption{Legacy (left) and BOSS (right) galaxies sky distributions,
   as available in DR10. In this study we use 801,945 galaxy spectra in the Legacy survey and 855,097 in the BOSS survey.}
 \label{fig:Galaxies_Sky_Distribution}
\end{figure*}

\begin{figure}
  \resizebox{84mm}{!}{\includegraphics{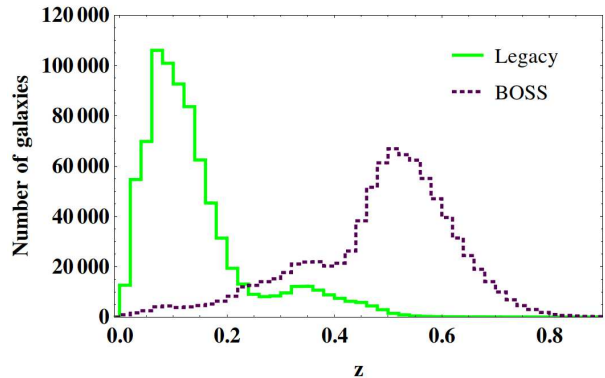}}
  \caption{Legacy (continuous, green) and BOSS (dashed, purple) galaxy redshift distributions.}
  \label{fig:Galaxies_Redshift_Distribution}
\end{figure}

The Legacy survey spectroscopic redshifts were obtained as part of
the SDSS-I and SDSS-II programmes \citep{York2000}, over an observing period of eight years,
shared with two additional surveys, the Sloan Extension for Galactic Understanding and
Exploration (SEGUE, for stars) and a Supernova survey. The Legacy Survey, originally
designed to investigate the large-scale structure of the universe is
composed of:
\begin{itemize}
  \item the \emph{Main} sample \citep{Strauss2002}, a
    magnitude-limited sample of galaxies with $r$-band Petrosian
    magnitudes $r<17.7$, and a
    median redshift of $z\sim 0.1$;
  \item and the \emph{Luminous Red Galaxies} (LRG) sample \citep{Eisenstein2001}, an
    approximately volume-limited sample up to $z\sim0.4$.
\end{itemize}
With a total sky coverage of 8,032 deg$^2$, as shown in
Fig.~\ref{fig:Galaxies_Sky_Distribution}, 
the Legacy Survey
includes over 930,000 unique galaxies with a spectroscopic redshift. 
Of those, we select the most reliable spectra with flags 
\texttt{ZWARNING} equal to 0 or 16, \texttt{plateQuality}
``good'' or ``marginal'', and \texttt{Z\_ERR} $< 0.0006$. Within the
redshift range in common with the cluster catalogues listed in
Sec.~\ref{sect:cluster_cats}, we find 801,945 galaxy spectra
useful for our purposes. Although the spectroscopic sample and sky coverage of
the Legacy sample have remained unchanged,
the imaging and the spectroscopic pipelines have been improved in subsequent SDSS data
releases. Thus, here we use the
Legacy survey spectra of the latest DR10 release \citep{Ahn2014}. The redshift
distribution of this sample of galaxies is shown in
Fig.~\ref{fig:Galaxies_Redshift_Distribution}. Most of these
galaxies are confined in the range $0<z<0.2$, with an extra contribution coming from LRGs at higher redshift that peaks at $z \sim 0.35$

The ongoing Baryon Oscillation Spectroscopic Survey (BOSS, \citealt{Dawson2013}) is part of
the six-year SDSS-III programme \citealt{Eisenstein2011}, which aims to obtain the
spectroscopic redshifts of 1.5 million luminous red galaxies out to
$z=0.7$, and the Lyman-$\alpha$ absorption lines of 160,000 quasars
in the range $2.2 < z < 3$, to measure the acoustic scale with
a precision of 1\% at redshifts $z=0.3$ and $z=0.55$. 
This latest publicly available set of data provides the
spectra of 859,322 unique galaxies over 6,373~deg$^2$
(in green in Fig.~\ref{fig:Galaxies_Sky_Distribution}).
As we did with the Legacy spectra, we select only the
most reliable galaxies, imposing \texttt{ZWARNING\_NOQSO=0} to be 0, and
removing objects with \texttt{plateQuality} set to ``bad'' or \texttt{Z\_ERR\_NOQSO} $>
0.0006$. This gives us a sample of 855,097 galaxies. As seen from the redshift distribution in
Fig.~\ref{fig:Galaxies_Redshift_Distribution}, most of BOSS galaxies lie above $z\sim0.4$.

\section{Gravitational Redshift and Other Internal Redshift Effects}
\label{sect:grav_redshift}

General relativistic time dilation means light emitted from within a gravitational potential is redshifted relative to a distant observer, in 
proportion to the potential depth in the weak field limit. This means we may expect centrally located BCG galaxies to be relatively more affected than the average cluster member \citep{Broadhurst2000}. This relative gravitational redshift (GR), $\Delta z_{\mathrm{GR}}$, is proportional to $\Delta \Phi / c^2 $, where $\Delta \Phi$ is the gravitational potential difference between the cluster galaxy and the cluster BCG. \citeauthor{Wojtak2011} claimed in 2011 to have measured for the first time this effect, using the GMBCG cluster catalogue and DR7 data. Analytic models \citep{Cappi1995} predict a gravitational redshift of the order of $c\, \Delta z_{\mathrm{GR}} \sim 10$\,km\,s$^{-1}$ for clusters with masses $\sim 10^{14} M_\odot$, consistent with \citeauthor{Wojtak2011} measurements, and as much as 300\,km\,s$^{-1}$ for clusters with masses $\sim 10^{16} M_\odot$.

Making use of N-body simulations in a $\Lambda$CDM universe, \citet{Kim2004} concluded that, assuming a redshift accuracy of 30\,km\,s$^{-1}$, over 5,000 clusters with masses above $\sim 5 \times 10^{13} M_\odot$ were needed in order to measure the gravitational redshift effect at the $2\sigma$ level. An important result of their study is that, above masses $\sim 10^{14} M_\odot$, the gravitational redshift signal is proportional to the cluster velocity dispersion, and hence, the number of clusters actually needed to detect the gravitational redshift signal does not depend on the mass of the clusters used. This is because the dispersion in the velocity difference between the BCG and the rest of the galaxies is found to increase with cluster mass in the simulations, adding to the inherent noise. They stress that a convincing detection would require sufficient data so that independent mass bins can be compared to examine the signature of gravitational redshift as a function of cluster mass.

Shortly after the \citeauthor{Wojtak2011} claim, \citet{Zhao2013} pointed out a potentially significant additional new blueshift deviation effect related to the special relativistic transverse doppler effect (TD) generated by random motion of the galaxies moving within the cluster potential. This additional shift $\Delta z_{\mathrm{TD}}$ is equal to $(\langle |\textbf{v}_{\mathrm{gal}}|^2 \rangle - |\textbf{v}_{\mathrm{BCG}}|^2)/2c^2$, opposite in sign to the GR shift, and of the same order of magnitude for clusters in virial equilibrium. In fact, for an spherical cluster in equilibrium, this yields $c\,\Delta z_{\mathrm{TD}}=(3\sigma^2_{\mathrm{obs}}-3\sigma^2_{\mathrm{BCG}})/2c$, where $\sigma_\mathrm{obs}$ is the observed line-of-sight velocity dispersion of the galaxies around the BCG, and $\sigma_\mathrm{BCG}$ is the velocity dispersion associated to BCGs, which is taken to be $\sigma_{\mathrm{BCG}} \sim \sigma_{\mathrm{obs}}/3$ by \citeauthor{Wojtak2011} and \citeauthor{Zhao2013}

More recently, \citet{Kaiser2013} has raised other significant corrections. 
As we are observing galaxies in our past light cone (LC), and due to the time it takes light to travel trough the cluster, we will see on average more galaxies moving away from us than toward us. This effect is compared by \citeauthor{Kaiser2013} to the one that ``causes a runner on a trail to meet more hikers coming toward her than going in the same direction''. This results in another shift of the distribution of galaxies around the BCGs equal to $c\,\Delta z_{\mathrm{LC}}=(\langle |\textbf{v}_{\mathrm{los\,gal}}|^2 \rangle - |\textbf{v}_{\mathrm{los\,BCG}}|^2)/c$, equal in sign to the TD effect, and of the same order of magnitude.

In addition to that, according to \citeauthor{Kaiser2013} we also have to deal with the fact that we are working with a magnitude-limited sample of galaxies: although the cluster catalogues, that are obtained using photometric data, are claimed to be volume complete up to a certain redshift limit $z \sim 0.4$, the sample of galaxies with measured redshifts is usually magnitude-limited, so that proper motion, changing the surface brightness (SB), or equivalently the apparent luminosity of galaxies due to the relativistic beaming effect, will bias the distribution of galaxies selected within clusters. For low velocities, this change in the luminosity is equal to $\Delta L/L=(3+\alpha(z))\,v_{\mathrm{los}}/c$, with $L$ the apparent luminosity of the galaxy, and $\alpha(z)$ the effective spectral index that takes into account the change in frequency and the resulting response of the photon count detector to this change. 

The modulation on the number of observable galaxies is thus obtained multiplying $\Delta L/L$ by the logarithmic derivative of the comoving density of observable objects $n_{\mathrm{obs}}$ above the luminosity limit $L_{\mathrm{lim}}(z)$, $\mathrm{d\ln}\,n_{\mathrm{obs}}[>L_{\mathrm{lim}}(z)]/\mathrm{d\ln}\,L$, which depends on the redshift distribution and the luminosity function associated to the galaxy survey considered, and its average on the redshift limits considered. This density modulation introduces a change on the observed distribution of galaxies equal to ${c\,\Delta z_{\mathrm{SB}}=-\langle (3+\alpha(z))\,\mathrm{d\ln}\,n/\mathrm{d\ln}\,L\rangle\,\langle |\textbf{v}_{\mathrm{los}}|^2 \rangle /c}$. In opposition to the others TD and LC dynamical effects, this new shift introduces a net blue-shift, also of the same order of magnitude.

Hence, the total redshift velocity difference $v_{los}$ between a galaxy and the central BCG is given by:
\begin{equation}
v_{los}=H(z)\,(d_{\mathrm{gal}}-d_{\mathrm{BCG}})+v_{pec}+c\,\Delta z \,,
\end{equation}
where $H(z)$ is the Hubble parameter, $d$ is the distance between the observer and the object in Mpc, $v_{pec}$ is the velocity due to the peculiar motion of the galaxy, and $\Delta z$ is the term arising from the combination of the previously mentioned distortions. From now on, we will refer to the combination of all these effects as an ``internal redshift''.

\subsection{Model}

We compute now the expected internal redshift, $\Delta z$, coming from the previously mentioned gravitational redshift (GR), transverse doppler (TD), past light cone (LC) and surface brightness (SB) effects for GMBCG, WHL12 and redMaPPer catalogues. The internal redshift that one would observe at a projected transverse distance $r_\perp$ from the center of a cluster halo with mass $M$ would be:
\begin{equation}
\Delta z = \Delta z_{\mathrm{GR}} + \Delta z_{\mathrm{TD}} + \Delta z_{\mathrm{LC}}\,+ \Delta z_{\mathrm{SB}}\,, 
\end{equation}
with:
\begin{equation}
\Delta z_{\mathrm{GR}}=\frac{-2}{c^2\,\Sigma(r_\perp)}\,\int_{r_\perp}^\infty \Delta \Phi (r) \frac{\rho_{\mathrm{NFW}}(r)\,r\,\mathrm{d}r}{\sqrt{r^2-r_\perp^2}}\,,
\end{equation}
\begin{equation}
\Delta z_{\mathrm{TD}}=\frac{1}{2\,c^2}\,(\langle |\textbf{v}_{\mathrm{gal}}|^2 \rangle - |\textbf{v}_{\mathrm{BCG}}|^2)\,,
\end{equation}
\begin{equation}
\Delta z_{\mathrm{LC}}=\frac{1}{c^2}\,(\langle |\textbf{v}_{\mathrm{los\,gal}}|^2 \rangle - |\textbf{v}_{\mathrm{los\,BCG}}|^2)\,,
\end{equation}
\begin{equation}
\Delta z_{\mathrm{SB}}=\frac{-\langle |\textbf{v}_{\mathrm{los\,gal}}|^2 \rangle}{c^2}\,\langle (3+\alpha(z))\,\frac{\mathrm{d\ln}\,n_{\mathrm{obs}}[>L_{\mathrm{lim}}(z)]}{\mathrm{d\ln}\,L}\rangle\,,
\end{equation}
where $\Sigma$ is the projected surface density of the NFW density profile $\rho_{\mathrm{NFW}}$ of a cluster halo with mass $M$, and $\Delta \Phi (r)$ is the potential energy difference between $r$ and the center of such halo. We use the prescription given by \citeauthor{Zhao2013} to compute $\langle|\textbf{v}_{\mathrm{gal}}|^2\rangle$ as a function of the potential via the isotropic Jeans equation:
\begin{equation}\label{eq:dispersion}
\langle|\textbf{v}_{\mathrm{gal}}|^2\rangle=3\,\sigma_{\mathrm{los}}^2=3\,\langle \sqrt{r^2-r_\perp^2}\,\frac{\partial \Phi}{\partial (\sqrt{r^2-r_\perp^2})} \rangle\,.
\end{equation}

The value of $\alpha(z)$ can be taken to be approximately 2, and, to compute $\mathrm{d}\ln n/\mathrm{d}\ln L$, we take $r<17.77$ as the apparent magnitude limit for our galaxies sample, and use the estimate of the luminosity function in the $^{0.1}$\textit{r}-band given by \citet{Montero2009} based on DR6 data, whose Schechter best fit parameters are: $\Phi_*=0.0093$, $M_*-5\log_{10} h=-20.71$ and $\alpha=-1.26$. It would be more accurate to use the specific luminosity function associated to galaxies belonging to the clusters considered, but it was shown by \citet{Hansen2009} that it doesn't differ much from the overall survey luminosity function, so we can use it as a good approximation. To calculate the average we use the Legacy galaxy distribution seen in Fig.~\ref{fig:Galaxies_Redshift_Distribution}:
\begin{equation}
\langle \mathrm{d}\ln n/\mathrm{d}\ln L \rangle=\frac{\int_{z_1}^{z_2}(\mathrm{d}\ln n/\mathrm{d}\ln L)\,(\mathrm{d}N/\mathrm{d}z)\, \mathrm{d}z}{\int_{z_1}^{z_2} (\mathrm{d}N/\mathrm{d}z)\, \mathrm{d}z}\,,
\end{equation}
where the lowest $z_1$ and highest $z_2$ redshift limits of integration are chosen according to the cluster sample considered in each case.

The velocity total shift observed from a stacked sample of cluster haloes would then be:
\begin{equation}
\Delta(r_\perp)=c\,\frac{\int_{M_1}^{M_2} \Delta z\,(M,r_\perp)\,\Sigma (r_\perp)\,(\mathrm{d}n(M)/\mathrm{d}M)\,\mathrm{d}M}{\int_{M_1}^{M_2} \Sigma (r_\perp)\,(\mathrm{d}n(M)/\mathrm{d}M)\,\mathrm{d}M }\,
\end{equation}
where we integrate between the mass range defined by the lowest $M_1$ and highest $M_2$ masses considered in each catalog, and the mass distribution in each case is given by $\mathrm{d}n(M)/\mathrm{d}M$, which is functionally fitted from the observed distribution of clusters, as in Fig.~\ref{fig:Clusters_Mass_Distribution}, but in this case considering only those that were not discarded in the process, that is, with sufficient nearby galaxies with spectroscopic measurements for a meaningful measurement. The model curves predict almost the same internal redshift for both GMBCG and WHL12 cluster samples, as the lowest mass and the mass distribution of clusters are almost identical for both catalogues. The main difference between these two catalogues, i.e., WHL12 ranging to higher masses, does not change the shape of the model curve too much as the contribution coming from high mass clusters is highly suppressed by the low values of the mass function distribution at these scales. The redMaPPer model curve, in the other hand, predicts a higher signal, which is consistent with the fact that redMaPPer minimum mass cutoff is much more conservative than GMBCG and WHL12 ones, resulting in a higher average cluster mass.

\subsection{Experimental Results}

\begin{figure*}
 \centering
 \resizebox{\textwidth}{!}{\includegraphics{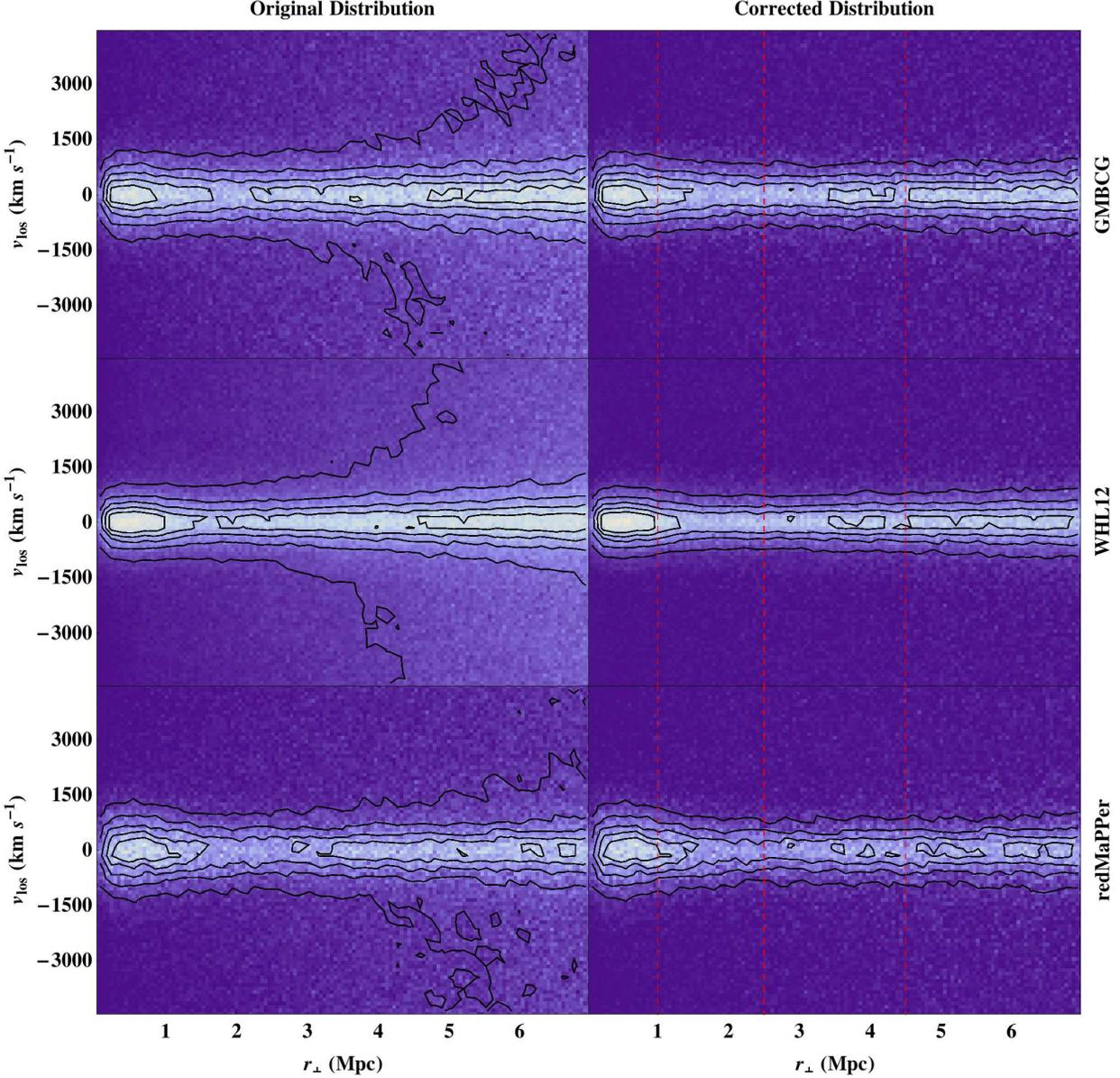}}
 \caption{GMBCG (top), WHL12 (middle) and redMaPPer (bottom) phase-space diagrams before (left) and after (right) removing statistically the foreground and background contribution of galaxies. Black contours represent iso-density regions. The asymmetry between the positive and negative $v_{los}$ region can be particularly clearly seen in the redMaPPer case. This difference disappears after the statistical interloper removal. We also plot as red dashed lines the boundaries at 1, 2.5 and 4.5\,Mpc that will determine the radial bins we will use in Sec.~\ref{sect:grav_redshift}. In these diagrams, the position of the BCG is fixed at $r_\perp=0$\,Mpc and $v_{los}= 0$\,km\,s$^{-1}$ by definition, and the density is determined by the number of galaxies with spectroscopic redshift measurements around them.}
\label{fig:phase_spaces}
\end{figure*}

In order to study the spatial distribution of galaxies around clusters, first, we carefully remove the identified BCGs of the cluster catalogues from the SDSS galaxy catalogues. Here we take into account the fact that, according to SDSS specifications, two galaxies are considered as the same object if they are closer than 3 arcsecs in the Legacy Survey case, and 2 arcsecs in the BOSS Survey case. Also, this will help us identify which of the BCGs have the best spectroscopic measurements, so, taking a conservative approach, we will only work with those BCGs identified in our ``high quality'' SDSS galaxy sample, discarding this way BCG redshift measurements obtained from ``bad'' plates. This leaves us with a total sample of 19,867 BCGs in the GMBCG catalog, 52,255 in the WHL12 case, and 10,197 in the redMaPPer one. We compute the projected transverse distance $r_\perp$ and the line-of-sight velocity $v_{los}=c\,(z_{\mathrm{gal}}-z_{\mathrm{BCG}})/(1+z_{\mathrm{BCG}})$ of all SDSS galaxies with respect to the BCGs, and keep those that lie within a separation of $r_\perp < 7$\,Mpc and ${| v_{los} | < 6{,}000}$\,km\,s$^{-1}$ from these. It should be noted that, as we are working mainly in a low redshift region, the impact of the cosmological parameters used is not significant. Stacking all the obtained pairs into one single phase-space diagram, we get the density distributions shown on the left handside of Fig.~\ref{fig:phase_spaces}.

To remove the contribution of foreground and background galaxies not gravitationally bound to clusters, we adopt an \textit{indirect} approximation, where galaxies not belonging to clusters are not identified individually in each cluster, as in the \textit{direct} method, but taken into account statistically once all the cluster information has been stacked into one single distribution of galaxies. See \cite{Wojtak2007} for a detailed study of different direct and indirect foreground and background galaxies removal techniques.

In our case, we apply the following procedure: first, we bin the whole phase-space distribution in bins of size 0.04\,Mpc $\times$ 50\,km\,s$^{-1}$. After that, we take all those bins lying in two stripes $4{,}500\,\mathrm{km\,s^{-1}} < | v_{los} | < 6{,}000 \,\mathrm{km\,s^{-1}}$, where we assume that all the galaxies there belong either to the pure foreground or to the pure background sample. Then, we fit a quadratic polynomial dependent of both $v_{los}$ and $r_\perp$ to the points in both stripes, and use the interpolated background model to correct the ``inner'' phase-space region bins. We use a function that depends not only on $r_\perp$, but also on $v_{los}$; this is because at high redshifts, and due to observational selection, we may have more spectroscopic measurements of those galaxies that are closer to us with respect to the BCG (i.e., have a negative $v_{los}$), than further away (positive $v_{los}$). The background-corrected phase-space diagrams for the three cluster catalogues can be seen on the right handside of Fig.~\ref{fig:phase_spaces}.

In Fig.~\ref{fig:phase_spaces} we can spot two clearly distinguishable regions: the \textit{inner}, dynamically relaxed, region of the cluster at $r_\perp < 1.5 - 2$\,Mpc, where iso-density contours are closed, and, at larger radius, the \textit{outer} radial infall region, highly compressed along the line of sight. This characteristic trumpet-shaped phase-space distribution is applied as a ``caustic method'' \citep{Diaferio1999} related to the escape velocity, to infer cluster mass profiles dynamically, where many redshifts of cluster members can define the caustic location. See \cite{Zu2013} and \cite{Lam2013} for recent developments on the field.

Following the method described in \citet{Wojtak2011}, we split the background-corrected phase-space diagram into different transverse distance bins, and measure the velocity distribution within these bins. In order to fit this distribution and measure any possible deviation from $\langle v_{los}\rangle = 0$, we adopt the double Gaussian function form: $A\,\mathrm{exp}((v_{los}-\Delta)^2/2 \sigma_A^2)+B\,\mathrm{exp}((v_{los}-\Delta)^2/2 \sigma_B^2)$, where both Gaussians, each with different amplitude and variance, share the same mean velocity $\Delta$. We present now the results obtained for each of the catalogues used in our analysis.

\begin{figure*}
 \centering
 \resizebox{\textwidth}{!}{\includegraphics{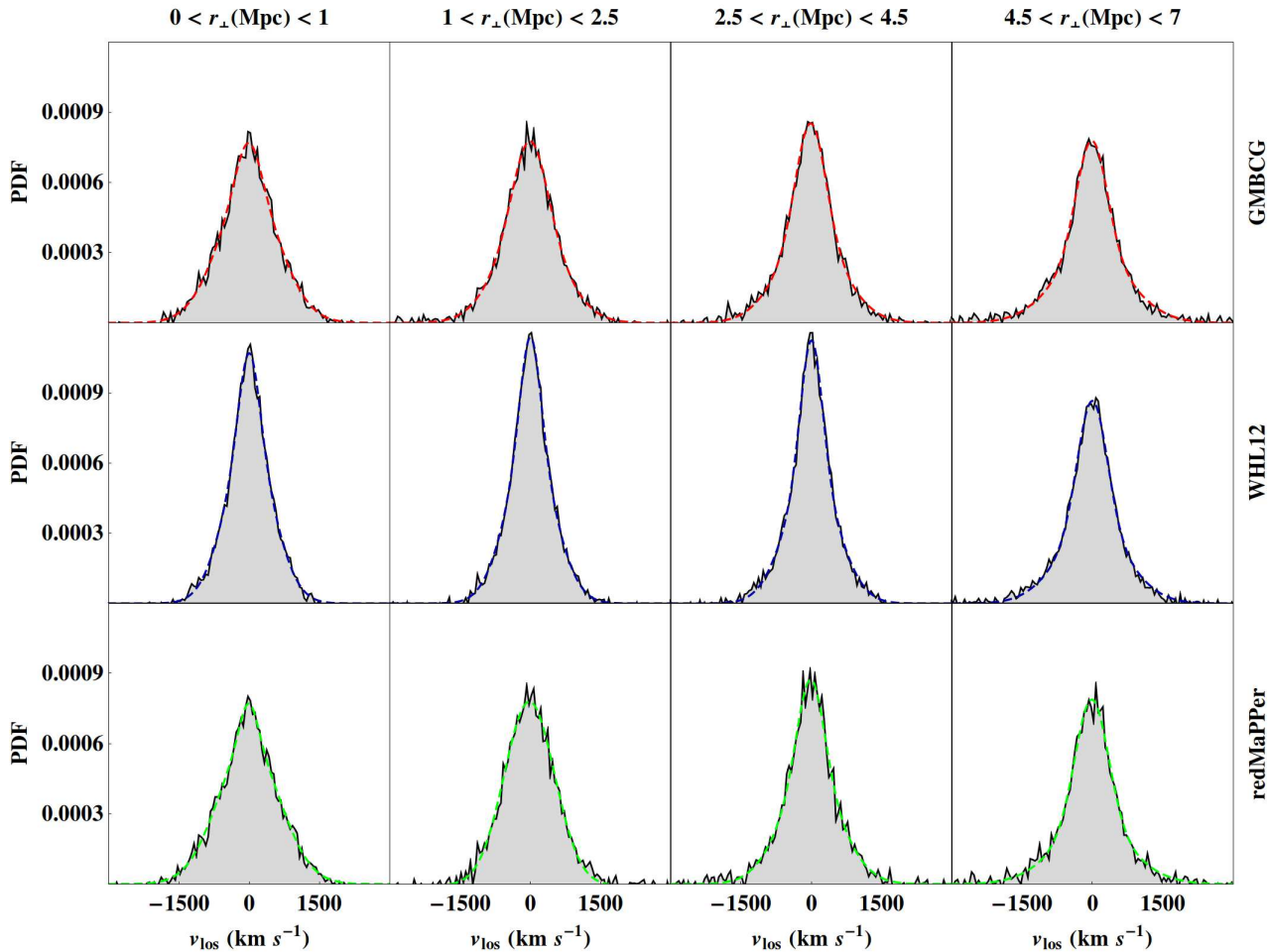}}
 \caption{GMBCG (top), WHL12 (middle) and redMaPPer (bottom) velocity distributions for 4 different radial bins, after the background and foreground contribution of galaxies has been removed. In top of these distributions, as dashed curves, the double Gaussian functional fit.}
\label{fig:velocity_ditribution}
\end{figure*}

\subsubsection{GMBCG Catalog}

In order to work only with the most reliable data, we decide to use only those clusters with 6 or more galaxies with spectroscopic redshift measurement in the previously defined phase-space region (7\,Mpc and $\pm4{,}500$\,km\,s$^{-1}$), that are located in the redshift range $0.1<z<0.4$ (where the catalogue is claimed to be complete), and with a richness greater than 10, which corresponds to $M_{200m} \sim 0.75 \times 10^{14}\,\mathrm{M}_\odot\,h^{-1}$. From the initial 19,867 clusters contained in the spectroscopic catalog, this leaves us with a sample of 4,278 objects, with mean richness 18 (corresponding to $M_{200m} \sim 1.5 \times 10^{14}\,\mathrm{M}_\odot\,h^{-1}$), and mean redshift $z\approx 0.22$. We display the velocity distribution and the resulting fits in the top part of Fig.~\ref{fig:velocity_ditribution}. The values of $\Delta$ obtained from the fits are displayed in Fig.~\ref{fig:grav_shift_radial}. These values are negative for all the radial bins considered, and seem to be consistent with the model proposed when considering all the effects described above, for which the prediction is a nearly flat profile of $\Delta \sim -10$\,km\,s$^{-1}$ at radius beyond $r_\perp>0.5$\,Mpc, from the central BCG position. Our measurements are compatible with those obtained by \citeauthor{Wojtak2011}, the difference between them coming from the different radial binning used.

We also divide the data into different mass bins in order to test how
this measurement may change with cluster mass. As before, we select only
those clusters with 6 spectroscopic BCG-galaxy pairs or more. Then, we
divide the resulting sample into 3 different mass subsamples. For each
of these subsamples we would like to measure the integrated signal up to
a certain radius $r_\perp$, but, as we expect cluster size to increase with
richness, a galaxy at, say, a distance of 0.5\,Mpc from a high richness
cluster's BCG would be deeper in the gravitational potential than a
galaxy 0.5\,Mpc away from a low richness cluster's BCG. So, in order to
make the measurements more comparable, we convert projected $r_\perp$ radial distances from
the BCGs into $r_{200c}$ units, i.e., we rescale the comoving
transverse distances of the galaxies that belong to a particular cluster 
using the $r_\mathrm{200c}$ estimate of that cluster,
obtained assuming a NFW halo density profile, and using the
concentration-mass relations provided by \cite{Bhattacharya2013}, as
explained before in Sec.~\ref{sect:data}. We then measure, for each
of the mass subsamples, the integrated signal of $\Delta$ up to
7\,$r_{200c}$, and the resulting values obtained are displayed in
Fig.~\ref{fig:grav_shift_stacked}. The first and the second mass
subsamples, with average masses of $\sim$ $0.8 \times
10^{14}\,\mathrm{M}_\odot\,h^{-1}$ and $1.4 \times
10^{14}\,\mathrm{M}_\odot\,h^{-1}$, show values of $\Delta$ equal to
$-11.2 \pm 3.2$\,km\,s$^{-1}$ and $-7.6 \pm 4.0$\,km\,s$^{-1}$,
respectively. The third mass subsample, with a higher average mass of
$\sim$ $3.6 \times 10^{14}\,\mathrm{M}_\odot\,h^{-1}$, gives a value of
$\Delta = -16.2 \pm 10.8$\,km\,s$^{-1}$. The error in this last
measurement is such that it seems inappropriate to claim an observed
signal dependence with increasing mass, despite the seemingly detection of a negative internal redshift signal for the three mass subsamples taken together.

\subsubsection{WHL12 Catalog}

Taking the same conservative approach as above, we discard all those clusters with less than 6 spectroscopic BCG-galaxy pairs in our phase-space defined region. On the other hand, although the WHL12 is claimed to be complete over a wider cosmological redshift range than GMBCG, we decide to adopt the same limited range as the GMBCG catalogue used above, $0.1<z<0.4$, in order to reduce the potential for any systematic differences between both measurements, making tier comparison easier to interpret. These limitations leave us with a sample of 12,661 clusters, with a mean richness of 23 (corresponding to $M_{200m} \sim 1.4 \times 10^{14}\,\mathrm{M}_\odot\,h^{-1}$), and a mean redshift of $z\approx 0.19$. The resulting velocity distribution and fits are displayed in the middle part of Fig.~\ref{fig:velocity_ditribution} for the 4 different radial bins used, and the fitted values of $\Delta$ are shown in Fig.~\ref{fig:grav_shift_radial}. As we can observe from the figure, the measured signal deviates completely from the proposed model: the first and fourth radial bins, centered at 1\,Mpc and 5.75\,Mpc, show values of $\Delta$ consistent with zero. Even worse, the second and third radial bins, centered at 1.75\,Mpc and 3.5\,Mpc, display positive values of $\Delta \sim + 5$\,km\,s$^{-1}$.

The number of clusters contained in the catalogue is large enough as to split it into different mass bins and still have enough number of objects to have a decent signal-to-noise ratio, so we proceed now to do it in order to test the reliability of this detection. In this case we divide those clusters with more than 5 galaxies with spectroscopic redshifts into 5 different mass subsamples. As before, we measure, for each of these mass subsamples, the integrated signal of $\Delta$ up to 7\,$r_{200c}$, where here we use the estimation of $r_\mathrm{200c}$ provided by the WHL12 cluster finder algorithm, a more direct indicator of the size and concentration of each cluster. The resulting values obtained are displayed in Fig.~\ref{fig:grav_shift_stacked}. In this case, the results obtained seem to be more illustrative than in the GMBCG case. The $\Delta$ value obtained from the first mass subsample, with an average $M_{200m} \sim 0.8 \times 10^{14}\,\mathrm{M}_\odot\,h^{-1}$, seems to be in agreement with the model prediction, but the signal obtained is very weak, compatible with zero at the $2\sigma$ level. The second and third mass subsamples, with average masses around 1.1 and $1.6 \times 10^{14}\,\mathrm{M}_\odot\,h^{-1}$, show positive values of $\Delta$. However, the fourth and fifth mass subsamples, whose average masses are $2.4 \times 10^{14}\,\mathrm{M}_\odot\,h^{-1}$ and $4.6 \times 10^{14}\,\mathrm{M}_\odot\,h^{-1}$ respectively, with values of $\Delta$ equal to $-17.2 \pm 7.2$\,km\,s$^{-1}$ and $-22.2 \pm 5.4$\,km\,s$^{-1}$, indicate a trend of a larger negative signal for larger cluster masses, corresponding to what one would expect from the model. We may think of this as a result of the cluster finding algorithm being more efficient in the task of identifying real clusters and their corresponding BCG for halo masses above $M_{200m} \sim 2 \times 10^{14}\,\mathrm{M}_\odot \, h^{-1}$, or the noise introduced by substructure and cluster mergers being less important for massive, relaxed clusters. In any case, it is clear that the positive values obtained in the radial global measurement of $\Delta$ are explained by the fact that the clusters in the WHL12 catalogue residing in this less massive region dominates over the more massive and ``reliable'' ones, for the mass distribution shown in Fig.~\ref{fig:Clusters_Mass_Distribution} indicates. The difference between GMBCG and WHL12 measurements may reside precisely in the fact that, as GMBCG algorithm is optimized to identify red sequence clusters and WHL12 relies only on galaxy FoF counting for their detection. The former method may contain a higher percentage of concentrated clusters, with a higher degree of virialization resulting in concordance between the measurement of internal redshift effects and the model for which virialization assumed.

\subsubsection{redMaPPer Catalog}

\begin{figure*}
\resizebox{100mm}{!}{\includegraphics{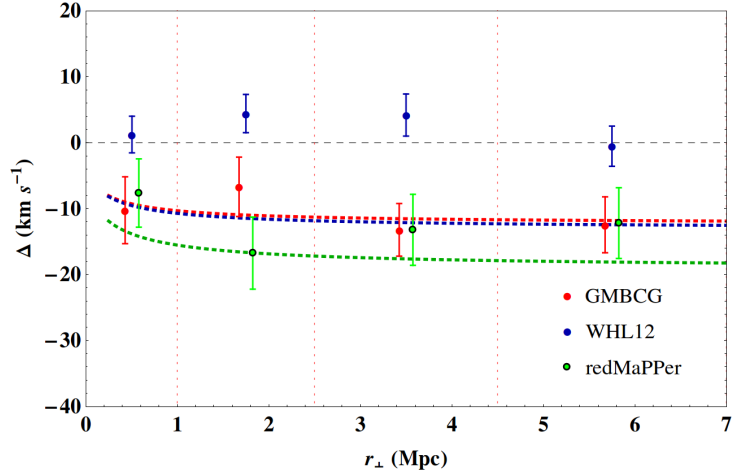}}
\caption{GMBCG (red), WHL12 (blue) and redMaPPer (green) $\Delta$ points for 4 different projected radial bins. The $r_\perp$ boundaries of these projected radial bins, marked as dashed red vertical lines, are: 0, 1, 2.5, 4.5 and 7\,Mpc. Dotted curves represent predictions from model. GMBCG and WHL12 model curves are almost identical, as the mass distribution of the clusters contained in these catalogues is very similar. On the other hand, redMaPPer clusters are, on average, much more massive, leading to an expected stronger effect.}
\label{fig:grav_shift_radial}
\end{figure*}

In the redMaPPer catalog, we also restrict the sample to those clusters in the $0.1<z<0.4$ redshift range and with 6 or more galaxies with spectroscopic redshift measurements, reducing the number of useful clusters from 10,197 to only 3,372, these having a mean richness of 35 (corresponding to $M_{200m} \sim 3.2 \times 10^{14}\,\mathrm{M}_\odot\,h^{-1}$, the double than in the two previously considered catalogues), and a mean redshift of $z\approx 0.23$. The velocity distribution with the corresponding fits and the resulting values of $\Delta$ obtained from them are shown in Figs.~\ref{fig:velocity_ditribution}~and~\ref{fig:grav_shift_radial} respectively. Although the amplitude of the signal is expected to be higher for this cluster sample, apart from the second radial bin centered at $r_\perp = 1.75$\,Mpc, with $\Delta = -16.7 \pm 5.5$\,km\,s$^{-1}$, all the other radial measurements of $\Delta$ do not deviate more than $\pm 2$\,km\,s$^{-1}$ from the measurements obtained using the GMBCG catalogue. Even when all the $\Delta$ measured points remain negative, there is no clear evidence for a stronger internal redshift signal compared to the one provided by GMBCG catalogue.

\begin{figure*}
\resizebox{100mm}{!}{\includegraphics{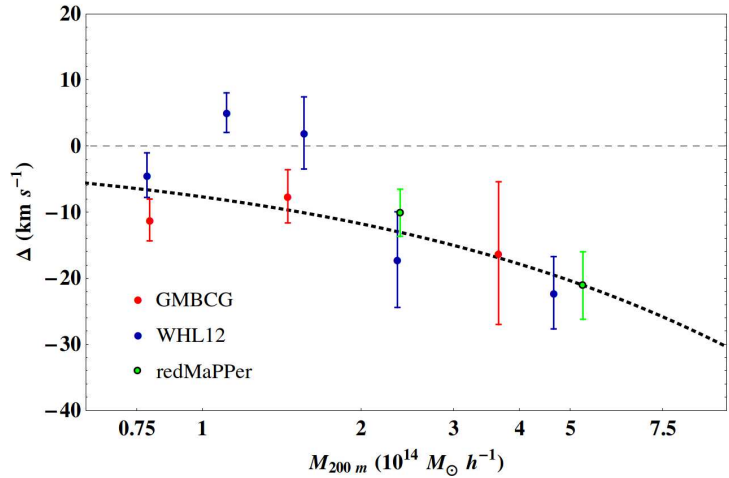}}
\caption{Measurements of the integrated signal of $\Delta$ to a distance of 7\,$r_{200c}$, for 3 different mass bins in the GMBCG cluster sample (red), 5 bins in the WHL12 one (blue), and 2 in the redMaPPer case (green). Dotted curve represents model prediction.}
\label{fig:grav_shift_stacked}
\end{figure*}

Now, as the number of clusters is relatively small, we measure the
integrated signal of $\Delta$ up to 7\,$r_{200c}$ for only two mass
subsamples of clusters. In this case, we use, as in the GMBCG case, the
$r_{200c}$ estimates obtained assuming a NFW density profile for
the clusters considered. 
The resulting measurements for the two mass bins,
$\Delta=-10.1\pm 3.5$\,km\,s$^{-1}$ and $-21.1\pm 5.1$\,km\,s$^{-1}$ at
$M_{200m}=2.4\times 10^{14}M_\odot\,h^{-1}$ and 
$5.3\times 10^{14} M_\odot\,h^{-1}$, respectively,
are displayed in Fig.~\ref{fig:grav_shift_stacked}. 
This comparison shows that the measured $\Delta$ appears to be more
negative for the high-mass sample than for the low-mass one at 1.8$\sigma$ significance. It is also reassuring that these measurements closely
follow the model prediction.


\section{Redshift Enhancement}
\label{sect:redshift_enhancement}

In this section we describe the measurement of
the redshift enhancement of background
galaxies behind the SDSS clusters due to lens magnification. 
This may help elucidate further the results we have found above for
the gravitational redshift. We are interested to see to what extent the
three cluster samples provide a consistent level of projected mass as
determined by a completely independent mass estimate generated by the
effect of gravitational lensing.

Lens magnification is caused in this case by a foreground cluster, which
acts as a gravitational lens with the lensing shear $\gamma(r)$ and
convergence $\kappa (r)= \Sigma(r) /
\Sigma_{\mathrm{crit}}$, where $\Sigma(r)$ is the projected mass density
of the lens in units of the critical surface density for lensing:
\begin{equation}
\Sigma_{\mathrm{crit}}=\frac{c^2}{4 \pi G}\frac{D_s}{D_l D_{ls}}\,,
\end{equation}
with $D_s$, $D_l$ and $D_{ls}$ referring to observer-source,
observer-lens and source-lens angular-diameter distances,
respectively. 

The magnification caused by the lens is given by:
\begin{equation}
 \mu = \frac{1}{(1-\kappa)^2-|\gamma|^2}\,,
\end{equation}
which distorts the background region in two ways: i) as gravitational
lensing preserves surface brightness, the flux from a source is
amplified as the lens increases the solid angle under which such source
appears. This implies that the luminosity limit of a survey is increased
by a factor $L_{\mathrm{lim}}/\mu$ in the lens region, resulting in a
higher surface density of observed background objects due to the ones
which could not had been seen otherwise. ii) The sky area behind the
foreground lens is expanded, so that the surface density of objects
decreases as the effective cross-section behind the clusters becomes
smaller. The combination of this two effects and the resulting
difference on the number of lensed sourced detected is known as
\textit{magnification bias} \citep{Broadhurst1995}. 

Thus, if the observed apparent luminosity of a lensed source is given by
 $L_{\mathrm{obs}}=\mu L_\mathrm{0}$, the observed number of objects
 with luminosities bigger than $L_{\mathrm{lim}}$ is given by: 
\begin{equation}
 n_{\mathrm{obs}}[>L_{\mathrm{lim}}(z)]=\frac{1}{\mu}\,n_\mathrm{0}[>L_{\mathrm{lim}}(z)/\mu]\,,
\end{equation}
where the $1/\mu$ factor comes from the dilation of the sky solid angle. 
Hence, in the case where $n_{\mathrm{0}}[>L(z)]\propto L(z)^{-\beta}$,
the previous equation simplifies to: 
\begin{equation}\label{eq:lensedVSunlensed}
 n_{\mathrm{obs}}(z)=\mu^{\beta(z,L) -1}\,n_\mathrm{0}(z)\,,
\end{equation}
where $\beta$ is the logarithmic slope of the luminosity function $\Phi$ evaluated at $L$:
\begin{equation}\label{eq:betafunction}
 \beta(z,L)= - \left. \frac{\mathrm{d}\ln\,\Phi(z,L')}{\mathrm{d}\ln\,L'} \right|_L\,.
\end{equation}
Taking into account that the number density of objects
$n_{\mathrm{obs}}(z)$ depends on redshift, the average redshift of the
background lensed sources is given by:
\begin{equation}
 \overline{z}_{\mathrm{back}}=\frac{\int n_{\mathrm{obs}}(z)\, z\, \mathrm{d}z}{\int n_{\mathrm{obs}}(z) \, \mathrm{d}z}\,,
\end{equation}
which, if $\beta(z,L)$ is greater than unity, is higher than the average
redshift in the absence of gravitational lenses.

\subsection{Model}
\label{sect:redshift_enhancement-model}

In order to model the expected redshift enhancement signal produced
by an ensemble of clusters, 
we first calculate the effect of magnification on the unlensed redshift distribution
$n_\mathrm{0}(z)$ of background sources using Eq.~(\ref{eq:lensedVSunlensed}).  
To compute the magnification $\mu$ as a function of mass and distance
from the cluster center, 
as done throughout the paper (Sec.~\ref{sect:data}),
we adopt the NFW density profile
with the concentration-mass relations
provided by \cite{Bhattacharya2013}, which are favored by 
recent cluster lensing
observations \citep{Okabe2013,Covone2014,Umetsu2014,Merten2014}. We employ 
the projected NFW functionals given by \cite{Wright2000}, which provide a 
good description of the projected total matter distribution of
cluster-sized haloes out to approximately twice the virial radius, beyond
which the two-halo term cannot be ignored
\citep{Oguri2011,Umetsu2014}. As we shall see, however, this projected
NFW model is sufficient to describe the data with the current
sensitivity.

As for the luminosity function $\Phi$ of the source galaxies, from
which we compute the logarithmic slope $\beta$
(Eq.~(\ref{eq:betafunction})), 
we follow CBU13 and use the Schechter
parametrization of the $V$-band luminosity function given by \cite{Ilbert2005}, 
obtained using VIMOS VLT Deep Survey \citep{LeFevre2005} data, and
adopt the redshift evolution from \cite{Faber2007}:
${M_*=-22.27-1.23 \times (z-0.5)}$ and $\alpha=-1.35$. The advantage of
using this particular survey, although its small 1\,deg$^2$ survey area,
resides in that it is much deeper (${0.2<z<2.0}$, $i < 24$) than the background
galaxy sample we are using (${0.45<z<0.9}$), so that the logarithmic
slope of the luminosity function as a function of redshift is very well
described in the range of redshift and magnitude we are interested
in.

Finally, the limiting luminosity used to evaluate
$\mathrm{d}\ln\,\Phi(z,L')/\mathrm{d}\ln\,L'$, is given by: 
\begin{equation}
-2.5 \log_{10} L(z)=i_{\mathrm{AB}}-5 \log_{10} \frac{d_L (z)}{10\,\mathrm{pc}}-K(z)\,,
\end{equation}
with $i_{\mathrm{AB}}=19.9$ the limiting magnitude of the BOSS survey, and $K(z)$ the $K$-correction:
\begin{equation}
K(z)=2.5\,(1+z)+2.5\log_{10} \left ( \frac{L(\lambda_e)}{L(\lambda_0)} \right )\,,
\end{equation}
where the second term can be neglected as the
$V$-band rest-frame flux falls in the $i$-band at $z \sim 0.5$.

\subsection{Experimental Results}

Observationally, the redshift enhancement $\delta_z(r)$ of background
galaxies is defined as:
\begin{equation}
  \delta_z(r) \equiv \frac{\overline{z}(r)-\overline{z}_{\mathrm{total}}}{\overline{z}_{\mathrm{total}}}\,,
\end{equation}
where $\overline{z}_{\mathrm{total}}$ is the average redshift of the
unlensed $N_{\mathrm{back}}$ background BOSS galaxies:
\begin{equation}
  \overline{z}_{\mathrm{total}}=\frac{1}{N_{\mathrm{back}}}\sum \limits_{i=1}^{N_{\mathrm{back}}}z_i\,,
\end{equation}
and $\overline{z}(r)$ the average redshift of the lensed $n(r)$ background
galaxies inside a radial bin at a physical distance $r$ from the cluster BCG:
\begin{equation}
  \overline{z}(r)=\frac{1}{n(r)}\sum \limits_{i=1}^{n(r)}z_i\,.
\end{equation}

A redshift enhancement signal at a significance of $4\sigma$
was first detected by CBU13, who used
five different cluster catalogues and a total of 316,220 
background BOSS galaxies from 
an earlier data release (DR9). Compared to CBU13, here we use over a
factor of two increase in the number of background galaxies
(855,097 in total), however we restrict our analysis to 
those clusters with a BCG spectroscopic redshift to ease the
comparison with the gravitational redshift measurements.
We note that the more background galaxies somehow compensate
the fewer clusters used in the analysis, so that the signal-to-noise
ratio is similar to CBU13.

We measure the redshift enhancement signal as a function
of radius $r$ from the BCG for the full cluster sample $\delta_z(r)$,
and the radially integrated redshift enhancement as a function of mass
(assuming a richness-mass relation) $\delta_z(M_{200m})$.
We repeat the measurements for each of the three cluster catalogues described in
Sec.~\ref{sect:cluster_cats}.

To estimate the errors on our measurements we
generate 500 catalogues with 25,000 random objects each, 
distributed inside the BOSS angular footprint, 
and following the same redshift distribution as
the cluster catalogue of interest. Then, using the same radial or mass
binning, we measure $\delta_z$ in the exact same way as for the real
background galaxy sample, and define the error bars as 
the standard deviation of the 500 signals. We also compute the full
covariance matrices to account for the re-use of cluster-background
galaxy pairs in the stacked signal, when computing the significance.
As pointed out in more details by CBU13, here the level of systematic
is negligible compared to statistical errors.

\subsubsection{Radial Redshift Enhancement}

We measure $\delta_z(r)$ in seven logarithmically spaced radial bins
in the range $0.04 < r < 15$~Mpc.  To compare these results with our
gravitational redshift results (see Sec.~\ref{sect:grav_redshift}),
we consider only those clusters for which the BCG has a spectroscopic
redshift in the range $0.1<z<0.4$. To ensure a significant gap between
the cluster lenses and background galaxies and avoid physically
associated pairs, we only use the BOSS galaxies with a spectroscopic
redshift larger than $z=0.45$.

\begin{figure*}
  \resizebox{100mm}{!}{\includegraphics{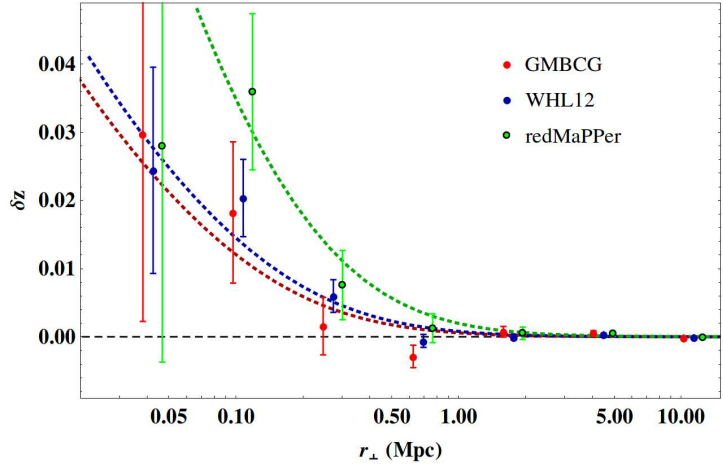}}
  \caption{Radial redshift enhancement signals for the GMBCG (red),
    WHL12 (blue) and redMaPPer (green) cluster catalogues. The dotted
    curves represent the model predictions for the three different
    considered cluster samples assuming a richness-mass relation. For visual
    clarity, the symbols for GMBCG and redMaPPer are horizontally shifted by $\mp 10\%$ with respect to WHL12.}
  \label{fig:radial_redshift_enhancement}
\end{figure*}

We show in Fig.~\ref{fig:radial_redshift_enhancement} the results
obtained for the GMBCG, WHL12, and redMaPPer cluster catalogues.
The model is computed as described in
Sec.~\ref{sect:redshift_enhancement-model},
 assuming a
richness-mass relation for individual cluster and summed over the
cluster mass distribution of each cluster sample, as shown in Fig.~\ref{fig:Clusters_Mass_Distribution}.
All measurements feature a $\delta_z(r)$ value in agreement with the
models within statistical errors. We note that the difference between
the redMaPPer model and the GMBCG/WHL12 models arises from the rather
different mass distributions. As seen in
Figure~\ref{fig:radial_redshift_enhancement}, the difference is most
significant at a scale of $\sim0.2-0.5$~Mpc.  The detection
significance of the redshift enhancement of background BOSS galaxies
behind clusters is calculated to be $2.8\sigma$, $4.7\sigma$ and
$3.9\sigma$ for the GMBCG, WHL12 and redMaPPer cluster catalogues,
respectively.

\subsubsection{Integrated Redshift Enhancement}

To study the mass dependence of this effect, as we have done in
Sec.~\ref{sect:grav_redshift}, we measure now the radially integrated
redshift enhancement signal in different mass bins.
To keep an approximately constant
signal-to-noise, we divide the GMBCG, WHL12 and redMaPPer
cluster samples into 3, 5 and 2 richness bins, respectively. 
We integrate $\delta_z(r)$ radially in the range $0.04 < r < 0.4$~Mpc,
where the signal-to-noise ratio is found to be highest.

\begin{figure*}
\resizebox{100mm}{!}{\includegraphics{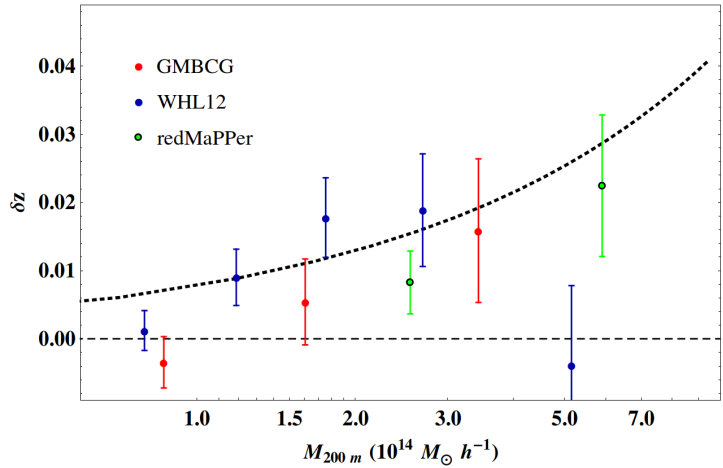}}
\caption{Radially integrated redshift enhancement signal
    in the range $0.04 < r < 0.4$, in 3 different richness bins in the case
    of the GMBCG cluster catalogue (red), 5 richness bins in the WHL12
    case (blue), and 2 richness bins in the redMaPPer case (green). Model
    prediction is shown as the dotted black curve.}
\label{fig:redshift_enhancement_stacked}
\end{figure*}

Results are displayed in
Figure~\ref{fig:redshift_enhancement_stacked}.
We report a clear tendency of an increasing value of $\delta_z$ with
increasing average cluster-sample mass, in qualitative agreement 
with the model. However we observe a $\sim1-2\sigma$ discrepancy at low mass ($M_{200m} < 1
\times 10^{14}\,\mathrm{M}_\odot\,h^{-1}$) for the GMBCG and WHL12 cluster
sub-samples, and a $\sim2-3\sigma$ discrepancy for the WHL12
sub-sample at high mass ($M_{200m} \sim 5 \times 10^{14}\,\mathrm{M}_\odot\,h^{-1}$).


\section{Discussion and Conclusions}
\label{sect:conclusions}

\subsection{Discussion}

Using the \textit{Legacy} sample of galaxies, comprising more than
800,000 galaxy spectra, we have detected two independent effects related
to the gravitational fields of galaxy clusters identified within this
survey. We have produced the phase-space distribution of galaxies around
the BCG spectroscopic positions provided by three major cluster
catalogues: GMBCG, with 20,119 clusters, WHL12, with 52,682 clusters, and
redMaPPer, with 13,128 clusters. 

We have then measured the internal redshift distortion $\Delta z$
associated to each cluster sample as a function of cluster radius. This distortion is identified as the deviation
from the BCG velocity of galaxies we associate with these clusters. We
have modeled this observational signature for each cluster survey
taking into account the combination of the gravitational redshift, the transverse-doppler, the past light cone, and the
survey-dependant surface brightness effects. The net gravitational
redshift effect that we derive is consistent with the expected cluster
richness-mass relation in the case of the GMBCG cluster sample, with
values of $\Delta$ around $-10$\,km\,s$^{-1}$. 

In the redMaPPer sample case, with a higher average cluster mass and a
lower number of clusters contained in it, the agreement between model
and observation is also good within the noise with a difference of at
most $\sim$ $+5$\,km\,s$^{-1}$ on average observed above the
expectation. In the WHL12 case, we observe an unexpected positive signal
ranging from $\sim0$\,km\,s$^{-1}$ to $\sim+5$\,km\,s$^{-1}$, in
complete disagreement with the model based on the richness-mass relation
proposed for this sample. 

If all our clusters were relaxed, had no sub-structure, and the number of spectroscopic measurements were proportional to the density of galaxies, each cluster would practically follow the stacked cluster distribution of Fig.~\ref{fig:phase_spaces}. However, such an ideal case is not realised due to the inevitable level of sub-structure, and from the observational selection effects and from algorithmic 
limitations in the definition of clusters and BCG galaxies. Even if BCG finder algorithms were perfect (in the sense of identifying the brightest most massive galaxy of each cluster), it has been shown by \citet{Skibba2011} that the implicit assumption that BCGs reside at the potential minimum is subject to an significant inherent variance leading to a biased measurement of the galaxy velocity dispersion arising from a difference between the measured position of the BCG and the real position of the cluster halo center \citep{Kim2004}. The underlying offset distribution between the dark matter projected center and the BCGs has been also studied by \cite{Zitrin2012} and \cite{Johnston2007}, being shown in the latter that the magnification signal is qualitatively less sensitive to the miscentering effect compared to the shear signal. 
%%%
In principle, stacking all the velocity distributions of galaxies around
BCGs into an effective distribution accounts for some of the previously
mentioned effects, and enables us to measure any statistical deviation
$\Delta$ from $\langle v_{\mathrm{gal}} \rangle=0$. This is what we
measured in Sec.~\ref{sect:grav_redshift}. However, if we look again
at the galaxy velocity distributions
(Fig.~\ref{fig:velocity_ditribution}) from which we measured $\Delta$,
we see that the velocity distributions obtained from GMBCG, WHL12 and
redMaPPer catalogues are different. A further analysis of these velocity
distributions shows that this difference holds for different ranges of
mass. 
%%%
In the ideal relaxed case, these profiles should follow
$\sigma_{\mathrm{obs}}^2=\sigma_{\mathrm{gal}}^2+\sigma_{\mathrm{BCG}}^2$,
relation from which $\sigma_{\mathrm{gal}}$ is obtained after assuming a
relation between the BCG motion and the velocity dispersion of satellite
galaxies, $\sigma_{\mathrm{BCG}}=\alpha\,
\sigma_{\mathrm{gal}}$. \citeauthor{Wojtak2011} and
\citeauthor{Zhao2013} consider $\alpha\simeq 0.3$, but it is pointed by
\citeauthor{Kaiser2013} that the frequent misidentification of BCGs as
central galaxies would lead to a higher value of $\alpha \sim 0.5$. 
%%%

Using the appropriate richness-mass relation for each
cluster sample, we analyzed the dependence of the integrated internal
redshift signal with mass, observing a clear correlation between the 
intensity of the signal $\Delta$ and the average mass of the sample,
especially in the range $M_{200m} > 2 \times
10^{14}\,\mathrm{M}_\odot\,h^{-1}$, where the measurements follow
particularly well the model. The positive radial $\Delta$ signal in the
WHL12 catalogue seems to mainly arise from lower mass clusters in the
range $M_{200m} < 2 \times 10^{14}\,\mathrm{M}_\odot\,h^{-1}$. 

We have also measured the level of magnification bias in each cluster
survey, using the latest DR10 BOSS 850,000 galaxy spectra, almost
tripling the number of galaxies used in the first measurement of this
effect by CBU13. We detect a clear radial redshift
enhancement of the background galaxies behind clusters in all three
surveys with a significance of $2.8\sigma$, $4.7\sigma$ and $3.9\sigma$
levels for GMBCG, WHL12 and redMaPPer cluster catalogues
respectively. 

Making use of the previously employed richness-mass
relations, we have also measured the integrated signal out to
$r_{\perp}=0.4$\,Mpc for different subsamples of clusters with different
average masses. After modeling this gravitational lensing feature using
projected NFW functionals for the clusters and luminosity functions based
on deep spectroscopic surveys, we find a generally good agreement
between theoretical predictions and observations for the three cluster
catalogues, with a clear increase of the mean redshift of background
sources at smaller decreasing projected clustocentric radius from the
BCG, and also an increasing redshift enhancement with increasing cluster
masses. The WHL12 catalogue follows less well the model for the low and
the high mass bins falling below the expected value, with discrepancies
of $2.1\sigma$ and $2.5\sigma$, respectively.

\subsection{Conclusions}

From a comparison of our internal redshift distortion and lensing
redshift enhancement measurements
for three major cluster samples defined from the SDSS survey,
we conclude that the WHL12 catalog, containing the largest number of
clusters, is anomalous in the sense that the net internal redshift effect
is found to be uniformly positive with radius at a level of
$+5$\,km\,s$^{-1}$ instead of negative with $\sim -20$\,km\,s$^{-1}$, as
expected given the claimed richnesses of these clusters. 
%%%

Examining the mass dependence of these results we find it is the
clusters with $M_{200m} < 2 \times 10^{14}\,\mathrm{M}_\odot \, h^{-1}$
that introduce the unexpected positive signal, as more massive
clusters produce a net redshift of $\sim -20$\,km\,s$^{-1}$, similar to GMBCG
and redMaPPer samples. Given the much higher number of clusters claimed
for the WHL12 sample compared to the other two catalogues, 
it could be that this positive signal arises from spurious
detection of clusters or from chance projection of less massive systems.
The internal redshift and lensing magnification signals have totally
different sensitivities to line-of-sight projection effects. It is very
likely that a higher degree of contamination due to projection effects
in this catalogue is responsible for the observed trends in both
measurements, as lensing measures the sum of the projected signal.
%%%

For the redMaPPer cluster catalogue, which has the smallest sample size due to its
conservative minimum richness cutoff, both measurements are shown to agree
well with respective predictions albeit the large statistical
uncertainties. 
It also exhibits the best performance in terms of the accuracy of cluster
mass estimates because the mass dependence of the signal predicted by models
is detected at the $1.8\sigma$ level.
Our promising measurements of the internal redshift and redshift
enhancement effects
obtained with the redMaPPer catalogue
bode well for future measurements
using upcoming large redshift surveys, such as DES, JPAS, eBOSS and EUCLID,
which will allow us to define large, clean samples of galaxy
clusters using such a robust algorithm.
%%%

Our analysis shows that internal redshift measurements 
are not simply limited by the statistical precision, namely the number of clusters
used, but are also sensitive to systematic effects that are not fully
understood.  In future work, we intend to study these systematics in
more detail utilizing phase space information to better account for the
inherent velocity dispersion of BCGs with respect to the mean cluster
velocity and other possible sources of systematics, such as the effects
of cluster miscentering, kinematic behavior of satellite galaxies in relaxed
and unrelaxed clusters, and substructures.

\hspace{0pt}\\ \textbf{Acknowledgements}\\

PJ and TJB thank the ASIAA for generous hospitality.
TJB is supported by IKERBASQUE, the Basque Foundation for Science.
RL is supported by the Spanish Ministry of Economy and Competitiveness through research projects FIS2010-15492 and Consolider EPI CSD2010-00064, and the University of the Basque Country UPV/EHU under program UFI 11/55.
PJ acknowledges financial support from the Basque Government grant BFI-2012-349.
TJB, RL and PJ are also supported by the Basque Government through research project GIC12/66.
KU acknowledges support from the Ministry of Science and
Technology of Taiwan through grants NSC 100-2112-M-001-008-MY3 and MOST
103-2112-M-001-030-MY3.

Funding for SDSS-III has been provided by the Alfred P. Sloan Foundation, the Participating Institutions, the National Science Foundation, and the U.S. Department of Energy Office of Science. The SDSS-III web site is http://www.sdss3.org/. SDSS-III is managed by the Astrophysical Research Consortium for the Participating Institutions of the SDSS-III Collaboration including the
University of Arizona,
the Brazilian Participation Group,
Brookhaven National Laboratory,
University of Cambridge,
Carnegie Mellon University,
University of Florida,
the French Participation Group,
the German Participation Group,
Harvard University,
the Instituto de Astrofisica de Canarias,
the Michigan State/Notre Dame/JINA Participation Group,
Johns Hopkins University,
Lawrence Berkeley National Laboratory,
Max Planck Institute for Astrophysics,
Max Planck Institute for Extraterrestrial Physics,
New Mexico State University,
New York University,
Ohio State University,
Pennsylvania State University,
University of Portsmouth,
Princeton University,
the Spanish Participation Group,
University of Tokyo,
University of Utah,
Vanderbilt University,
University of Virginia,
University of Washington,
and Yale University.

\label{lastpage}

\end{document}